\documentclass{JHEP3}
\usepackage{psfrag,epsf,amsmath,amssymb,graphicx,cite}
\usepackage{scalefnt}
\newcommand{\hpl}{{\abbrev HPL}}
\newcommand{\code}{\sc}
\newcommand{\abbrev}{\scalefont{.9}}

\newcommand{\api}{\frac{\alpha_s}{\pi}}
\newcommand{\eqn}[1]{Eq.\,(\ref{#1})}
\newcommand{\fig}[1]{Fig.\,\ref{#1}}

\newcommand{\sct}[1]{Sect.\,\ref{#1}}
\newcommand{\dd}{{\rm d}}

\newcommand{\order}[1]{{\cal O}(#1)}
\newcommand{\lo}{{\abbrev LO}}
\newcommand{\nlo}{{\abbrev NLO}}
\newcommand{\nnlo}{{\abbrev NNLO}}
\newcommand{\msbar}{\mbox{$\overline{\mbox{\abbrev MS}}$}}
\newcommand{\bld}[1]{\boldmath{$#1$}}

\newcommand{\qcd}{{\abbrev QCD}}
\newcommand{\sm}{{\abbrev SM}}
\newcommand{\mssm}{{\abbrev MSSM}}
\renewcommand{\Re}{{\rm Re}}

\newcommand{\ibe}{{\abbrev E{\scalefont{.9}\&}I}}
\newlength{\figwid}
\newcommand{\feynsl}[1]{
  \setbox0=\hbox{/} \setbox1=\hbox{$#1$}
  \dimen0=\wd0 \advance\dimen0 by -\wd1 \divide\dimen0 by 2
  \ifdim\wd0>\wd1 \lower.15ex
          \copy0\kern-\wd0\kern\dimen0\copy1\kern\dimen0
  \else \kern-\dimen0\lower.15ex
          \copy0\kern-\dimen0\kern-\wd1\copy1\fi}

\DeclareMathOperator{\Snp}{S}
\DeclareMathOperator{\Li}{Li}
\DeclareMathOperator{\HPL}{H}
\newcommand{\abs}[1]{{\ensuremath{\left| #1 \right|}}}

\title{Higgs production and decay: Analytic results at next-to-leading
  order QCD}
\author{Robert V. Harlander and Philipp Kant\\
  {\it Institut f\"ur Theoretische Teilchenphysik,
  Universit\"at Karlsruhe\\
  D-76128 Karlsruhe, Germany}\\
    E-mail: \email{robert.harlander@cern.ch},
\email{kantp@particle.uni-karlsruhe.de}\\
}

\preprint{\sf TTP05--18, SFB/CPP-05-49 --- September 2005}

\abstract{ The virtual two-loop corrections for Higgs production in
gluon fusion are calculated analytically in QCD for arbitrary Higgs and
quark masses.  Both scalar and pseudo-scalar Higgs bosons are
considered. The results are obtained by expanding the known
one-dimensional integral representation in terms of $m_H/m_q$, and
matching it with a suitably chosen ansatz of Harmonic
Polylogarithms. This ansatz is motivated by the known analytic result
for the Higgs decay rate into two photons. The method also allows us to
check this result and to extend it to the pseudo-scalar decay rate.}
\begin{document}
\setlength{\figwid}{7em}

\section{Introduction}
The gluon fusion process for Higgs production at a hadron collider has
been studied in great detail over the last few years (for a recent
review, see Ref.\,\cite{Djouadi:2005gi}).  It is well-known to be the
dominant mode in the Standard Model and also in most of the usually
considered supersymmetric parameter space.  The fact that the
next-to-leading order QCD
corrections~\cite{Dawson:1990zj,Djouadi:1991tk,Spira:1995rr} increase
the cross section by more than 70\% triggered more detailed studies of
higher order effects. In particular, the
\nnlo{}\cite{Harlander:2002wh,Anastasiou:2002yz,Harlander:2002vv,
Anastasiou:2002wq,Ravindran:2003um} and quite recently even the leading
threshold-enhanced {\abbrev N$^3$LO}~\cite{Moch:2004pa} corrections were
evaluated in the heavy-top limit, indicating a well-behaved perturbative
expansion of the total cross section.  Meanwhile, the \nnlo{} effects are
known also for differential quantities in terms of a partonic \nnlo{}
Monte Carlo program~\cite{Anastasiou:2004xq,Anastasiou:2005qj}, allowing
to simulate experimental cuts, for example.

In contrast to the \nnlo{} calculations which currently all rely on the
heavy-top limit, the inclusive \nlo{} effects were calculated for
arbitrary values of the Higgs boson mass and the mass of the quark that
mediates the gluon-Higgs
coupling~\cite{Graudenz:1992pv,Spira:1995rr,Spira:1997dg}. In fact, it
is this calculation that justifies the use of the heavy-top limit at
\nnlo{}, because it explicitely demonstrates the excellent quality of
this limit even at Higgs masses close to the quark threshold $m_H\approx
2m_q$ and beyond.  Probably the most important application of the
general $m_H/m_q$ dependence currently is supersymmetry, where bottom
quarks can contribute significantly to the gluon-Higgs coupling due to a
potential enhancement proportional to $\tan\beta$ of their Yukawa
coupling to Higgs bosons. For bottom quarks, an analogous
``heavy-quark'' approximation would certainly be very doubtful in this
context~\cite{Spira:1995rr,Kramer:1996iq,Spira:1997dg,Harlander:2003xy}.

Considering the importance of the full mass dependence, it is somewhat
surprising that the status of the \nlo{} calculation is still at the
level of more than ten years ago. By then, the result was obtained in
terms of a rather lengthy one-dimensional integral representation and
implemented in a {\abbrev FORTRAN} routine.  On the one hand, this makes
it rather difficult to import the result into other programs, of course.
On the other hand, it is practically impossible to further manipulate
the result.

The lack of an analytical result is also surprising in view of the great
technical progress since the original work of Ref.\,\cite{Spira:1995rr}.
In fact, the corresponding Feynman integrals belong to a class that
currently receives great attention due to its importance for
electro-weak precision observables (see, e.g.,
Refs.\,\cite{Bonciani:2004dz,Pozzorini:2004rm,Jantzen:2005az} and references
therein). It turns out indeed that all integrals needed for a
representation of the 2-loop virtual terms in closed form have been
evaluated in the literature.

In this paper we derive this analytic formula. Let us stress though that
we did not evaluate the corresponding Feynman diagrams; rather, we used
the integral representation given in Ref.\,\cite{Spira:1995rr} and
evaluated it analytically. The method we followed is rather
unconventional but not new. For the sake of brevity, we will refer to it
as {\it Expansion and Inversion (\ibe{})} in what follows.  It relies on
the identity theorem for power series: Two analytic functions are the
same if their Taylor series are the same. A more detailed description
of the method and its realization will be given in
Section\,\ref{sec::method}.

\section{Discussion of the method; calculation of the decay rates}%
\label{sec::method}

The idea behind our approach is that, if two physical processes
correspond to a similar set of Feynman diagrams (kinematics, mass
assignment), their cross sections should be described by a common set of
analytical functions. Thus, if one processes is known, one can establish
an ansatz for the other one by a linear combination of these functions,
with unknown coefficients.  In the \ibe{} method, one then evaluates the
power series of the unknown cross section in a certain limit and
compares it with the corresponding expansion of the ansatz. This leads
to a system of linear equations for the unknown coefficients which can
be solved uniquely if the depth of the expansion matches the number of
unknowns. In general, it is advisable to overdetermine the system in
order to confirm that the ansatz is complete.  The identity theorem for
power series ensures that the solution obtained in this way is indeed
the analytical result for the cross section.

It is important to realize that, while the intermediate power series
approximates the full result only within the radius of convergence, the
final result is valid for arbitrary values of the parameters. Thus, the
comparison of the final result with a numerical evaluation of the
original integral {\it outside} the radius of convergence provides one
of the most powerful checks on the calculation.

The main advantage of the \ibe{} approach is that, in most cases, the
power series of the cross section can be obtained in a rather simple
manner. A powerful tool for this goal is provided by asymptotic
expansions of Feynman diagrams (see
Refs.~\cite{Smirnov:1994tg,Smirnovbook,Harlander:1998dq} and references
therein). This method works directly at the level of Feynman integrals
and has been fully automated for the case of Euclidean external
momenta~\cite{Seidensticker:1999bb,Harlander:1997zb}.

Very often, however, one can derive one-dimensional integral
representations over finite integration regions. This can be achieved by
introducing Feynman parameters, for example, and performing all but one
integral analytically. If the interchange of integration and
differentiation is possible, the power series expansion can be performed
directly on the integrand, and the resulting integrals are in general
much simpler than the original ones.  

Another example where the integrations are over finite regions is given
by phase space integrals, and in fact, the \ibe{} method has been used
for the evaluation of the three-particle phase space integration in the
case of Higgs production and the Drell-Yan process, both at
\nnlo{}~\cite{Harlander:2002wh,Harlander:2002vv,
Kilgore:2002sk,Harlander:2003ai}.

In this paper, we apply the \ibe{} method to obtain analytic formulae
for the \nlo{} predictions of the Higgs decay rate into two photons, as
well as to the virtual two-loop corrections for Higgs production in
gluon fusion. We consider both scalar and pseudo-scalar Higgs bosons
such that our results are relevant also in supersymmetric scenarios or
other extensions of the \sm{}.

For all these quantities, a one-dimensional integral representation is
known~\cite{Spira:1995rr}. By interchanging differentiation and
integration, we can derive their power series in terms of $m_H/m_q$.  A
closed analytical result is only known for the \nlo{} decay rate of a
scalar Higgs boson into photons~\cite{Fleischer:2004vb}. We use it as
the main motivation of our ansatz in order to derive closed analytical
expressions for the other quantities as well.  It will be useful for the
rest of this paper to quote the explicit result at this point.

\subsection[Decay rate $H\to \gamma\gamma$]{%
  Decay rate \bld{H\to \gamma\gamma}}\label{sec::hgamgam}

The decay rate of a Higgs boson into two photons through \nlo{} can be
written as (see, e.g., Ref.\,\cite{Spira:1995rr})
\begin{equation}
	\Gamma(H\to\gamma\gamma) = 
	\frac{G_F \alpha^2 m_H^3}{128 \sqrt{2} \pi^3}
	\abs{
		\sum_l Q_l^2 A_l^H(\tau_l) 
		+ 3\sum_q Q_q^2 A_q^H(\tau_q) 
		+ A_W^H(\tau_W)}^2,
\label{eq::hgamgam}
\end{equation}
where $G_F$ is the Fermi constant, $\alpha$ is the electromagnetic
fine-structure constant, $m_H$ denotes the Higgs mass, and $Q_{q,l}$ the
electric charge of quark $q$ and lepton $l$ in units of the proton
charge.
The variables $\tau_{i}$ are defined as
\begin{equation}\label{def:tau}
	\tau_{i} := \frac{m_H^2}{4 m_i^2},
\end{equation}
where $m_i$ denotes the mass of particle $i$. Here and in what follows,
$m_q\equiv m_q(\mu)$ denotes the $\msbar$ quark mass renormalized at a
mass scale $\mu$.

\FIGURE{%
    \begin{tabular}{c}
      \includegraphics[width=.7\textwidth]{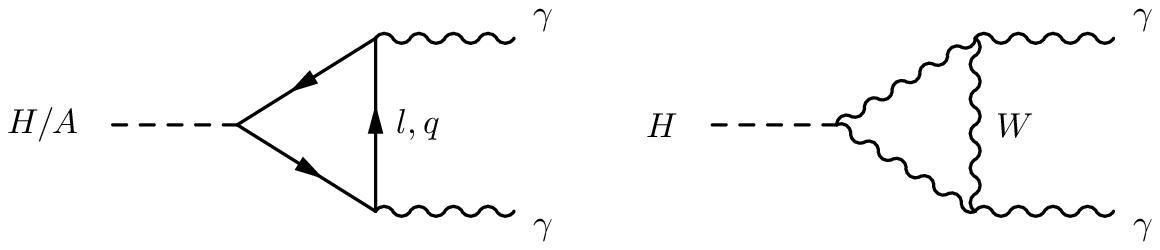}\\[0em]
      (a)\hspace{15em}(b)
    \end{tabular}
    \parbox{.9\textwidth}{
      \caption[]{\sloppy
	Feynman diagrams contributing to the decay rate for $H/A\to
	\gamma\gamma$ at \lo{}.
	\label{fig::hgamgam_dia.eps}
        }}
}

The amplitudes $A_l^H(\tau)$ and
$A_W^H(\tau)$ arise from closed lepton and $W$-boson loops,
respectively (cf.\,\fig{fig::hgamgam_dia.eps}), and do not receive \qcd{}
corrections. They are given by~\cite{Ellis:1975ap,Shifman:1979eb}
\begin{equation}
	\begin{split}
		A_l^{H}(\tau)
		& = \frac{2}{\tau^2} \bigl[ \tau + (\tau-1)f(\tau)
		\bigr]
		\equiv \frac{4}{3}\,F_0^H(\tau)
		\,,
		\\ 
		A_W^{H}(\tau) 
		& = -\frac{1}{\tau^2} \bigl[ 2\tau^2 + 3\tau 
			+ 3(2\tau-2) f(\tau)\bigr],  
\label{eq::alhawh}
	\end{split}
\end{equation}
where
\begin{equation}\label{eq:higgs:f}
	f(\tau) = \begin{cases} \arcsin^2\bigl( \sqrt{\tau} \bigr), &
		\tau \leq 1 \\ \displaystyle -\frac{1}{4}\biggl[
		\ln\frac{ 1 + \sqrt{1-\tau^{-1}} } { 1 -
		\sqrt{1-\tau^{-1}} } - i \pi \biggr]^2, & \tau > 1.
	\end{cases}
\end{equation}

\FIGURE{%
    \begin{tabular}{c}
      \includegraphics[width=.7\textwidth]{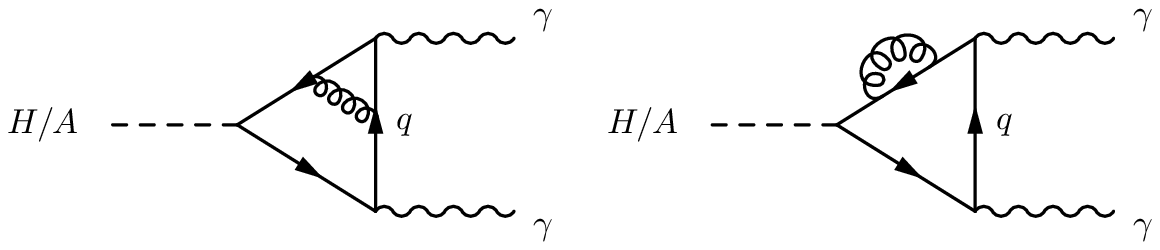}\\[0em]
      (a)\hspace{15em}(b)
    \end{tabular}
    \parbox{.9\textwidth}{
      \caption[]{\sloppy
	Sample Feynman diagrams contributing to the decay rate for
	$H/A\to \gamma\gamma$ at \nlo{}.
	\label{fig::hgamgam_nlo.eps}
        }}
}

$A_q^H(\tau)$ originates from the quark mediated photon-Higgs coupling,
\fig{fig::hgamgam_dia.eps}\,(a) and \fig{fig::hgamgam_nlo.eps}. Through
\nlo{} \qcd{}, one can write it as
\begin{equation}
\begin{split}
A_q^H(\tau) &= \frac{4}{3}\,F_0^{H}(\tau)\,\left[ 1 + \api\left(
  C_1^H(\tau) + C_2^H(\tau)\ln\frac{4\tau\mu^2}{m_H^2} \right)\right]\,,
\label{eq::aqh}
\end{split}
\end{equation}
with $F_0^{H}(\tau)$ from \eqn{eq::alhawh}.  $C_2^H$ follows directly
from the \nlo{} renormalization group equation
\begin{equation}
\begin{split}
\left(\mu^2\frac{\partial}{\partial\mu^2} +
2\,\api\,\tau\frac{\partial}{\partial \tau}\right) A_q^H(\tau) &=
\order{\alpha_s^2}\,.
\label{eq::rge}
\end{split}
\end{equation}
It reads
\begin{equation}
\begin{split}
F_0^H\,C_2^H &= \frac{3}{\tau^2}\left[
\tau + \left(\tau - 2\right)f(\tau) - (\tau-1)\tau\,f'(\tau)\right]\,.
\label{eq::c2h}
\end{split}
\end{equation}

The analytical expression for $C_1^H$ has been obtained in
Ref.\,\cite{Fleischer:2004vb}:\footnote{Eqs.\,(10) and (12) of
Ref.\,\cite{Fleischer:2004vb} contain typos; thanks to O.~Tarasov for
immediate confirmation. Note that in order to compare \eqn{eq::c1h} with
the formula in Ref.\,\cite{Fleischer:2004vb}, one needs to use the
identity $\Li_3(\theta^2)=4\bigl[\Li_3(\theta)+\Li_3(-\theta)\bigr]$.}

\begin{equation}
  \begin{split}
    F_0^H\,&C_1^H  =\\& 
    - \frac{ \theta \left(1+\theta+\theta^2+\theta^3\right) }
    {{\left(1-\theta\right)}^5} 
    \biggl[
       108\Li_4(\theta)
      + 144\Li_4(-\theta)
      - 64\Li_3(\theta)\ln\theta
      \\&\qquad
      - 64\Li_3(-\theta)\ln\theta
      + 14\Li_2(\theta)\ln^2\theta
      + 8\Li_2(-\theta)\ln^2\theta
      + \frac{1}{12}\ln^4\theta 
      \\&\qquad
      + 4\,\zeta_2\ln^2\theta
      + 16\,\zeta_3\ln\theta
      + 18\,\zeta_4
      \biggr] 
    \\& 
    +\frac{\theta {\left(1+\theta\right)}^2}{{\left(1-\theta\right)}^4}
    \biggl[
      -32\Li_3(-\theta)
      + 16\Li_2(-\theta)\ln\theta
      -4\,\zeta_2\ln\theta
      \biggr]
    \\&
    - \frac{4\, \theta \left(7-2\,\theta+7\,\theta^2\right) }
    {{\left(1-\theta\right)}^4}
    \Li_3(\theta)
    +\frac{8\, \theta \left(3-2\,\theta+3\,\theta^2\right) }
    {{\left(1-\theta\right)}^4}
    \Li_2(\theta)\ln\theta
    \\&
    +\frac{2\, \theta \left(5-6\,\theta+5\,\theta^2\right) }
    {{\left(1-\theta\right)}^4}
    \ln(1-\theta)\ln^2\theta
    +\frac{\theta \left(3+25\,\theta-7\,\theta^2+3\,\theta^3\right) }
    {3 {\left(1-\theta\right)}^5}
    \ln^3\theta
    \\&
    +\frac{4\, \theta \left(1-14\,\theta+\theta^2\right) }
    {{\left(1-\theta\right)}^4}
    \zeta_3
    +\frac{12\, \theta^2 }
    {{\left(1-\theta\right)}^4}
    \ln^2\theta
    -\frac{12\, \theta \left(1+\theta\right) }
    {{\left(1-\theta\right)}^3}
    \ln\theta
    -\frac{20\, \theta}
    {{\left(1-\theta\right)}^2}
    \,,
  \end{split}
  \label{eq::c1h}
\end{equation}

with Riemann's zeta function
\begin{equation}
\begin{split}
\zeta_n \equiv \zeta(n)\,,\qquad\mbox{i.e.}\quad
\zeta_2 = \frac{\pi^2}{6}\,,\quad
\zeta_3 = 1.20206\ldots\,,\quad
\zeta_4 = \frac{\pi^4}{90}\,,
\end{split}
\end{equation}
and
\begin{equation}
\begin{split}
  \theta\equiv\theta(\tau) = 
  \frac{\sqrt{1-\tau^{-1}}-1}{\sqrt{1-\tau^{-1}}+1}\,.
  \label{eq::thetadef}
\end{split}
\end{equation}
For analytic continuation, it is always understood that $\tau
\to \tau+i0$.

The products of logarithms and polylogarithms in \eqn{eq::c1h} can be
expressed in terms of Harmonic Polylogarithms~\cite{Remiddi:1999ew} of
the form $\HPL(\vec{n};\theta)$, where $\vec n$ is an $n$-tuple with
entries (``indices'') $\pm 1$ or 0.  One finds that $n\leq 4$, and that
at most one index is different from zero.

This suggests to construct our ansatz from Harmonic Polylogarithms of
this form, multiplied by rational functions
\begin{equation}
\begin{split}
R_{n,k}(\theta) &= \frac{P_n(\theta)}{(1-\theta)^k}\,,
\end{split}
\end{equation}
where $P_n(\theta)$ is a polynomial in $\theta$ of degree $n$ with
unknown coefficients. In order to solve the resulting system of linear
equations, one has to adjust the integer parameters $n,k$ such that a
suitable balance is obtained between the universality of the ansatz and
the depth of the power series expansion that is required to determine
the unknown coefficients.

As a warm-up, we may try to reproduce \eqn{eq::c1h} from the
one-dimensional integral representation of Ref.\,\cite{Spira:1995rr}
using our approach. To this aim, we expand the integrands of
$I_1,\ldots,I_5$, defined in Eqs.\,(A.9) to (A.13) of
Ref.\,\cite{Spira:1995rr},\footnote{ Thanks to M.~Spira for
clarification concerning some typos in the formulas of
Ref.\,\cite{Spira:1995rr}.}  around the limit $\tau=0$, keeping terms
through order $\tau^{100}$.

It is clear that due to the complexity of the integrands and the
required depth of the expansion we need to use efficient computer
algebra tools. We found that the {\code Taylor} package~\cite{taylor}
for {\code Reduce}~\cite{reduce} is particularly well suited for this
kind of operations. In most cases, the results obtained from {\code
Taylor} were checked against our own implementation of the relevant
power series in {\code Form}~\cite{Vermaseren:2000nd}.  The capabilities
of {\code Mathematica}~\cite{Mathematica}, on the other hand, are
clearly not suited for expanding expressions of this
complexity.\footnote{We remark that {\code Mathematica 5.1} even
produces a wrong result when expanding $\Li_2(1-x)$ around $x=0$ (also
for $\Li_3$ etc.); a bug report has been submitted and acknowledged.}

For illustration of the method, let us consider the simplest one of the
relevant integrals:
\begin{equation}\label{eq:i5}
	I_5 = \int_0^1 \frac{\dd x}{1-\rho x} \biggl\{
	\alpha_+ \ln\Bigl( 1 - \frac{x}{\alpha_+} \Bigr)
	+ \alpha_- \ln\Bigl( 1 - \frac{x}{\alpha_-} \Bigr)
	\biggr\} \ln\Bigl( \frac{1-\rho x(1-x)}{x} \Bigr),
\end{equation}
where
\begin{equation}
	\begin{split}
		\nonumber \rho & = 4 \tau = m_H^2/m_q^2, \\
		\alpha_\pm & = (1 \pm \sqrt{1-\tau^{-1}} )/2.
	\end{split}
\end{equation}
Expanding the integrand around $\tau=0$ leads to very simple integrations
in $x$,
\begin{equation}
	\begin{split}
		I_5 &= \int_0^1\dd x \Biggl\{
		2\,x\,\ln x 
		+	\tau \, \biggl[ 8\,x^2 - 8\,x^3 
			+ \ln x\Bigl(10\,x^2 - \frac{8}{3}\,x^3
			\Bigr)\biggr] 
		\\&
		+ \tau^2 \, \biggl[ 56\,x^3 - \frac{248}{3}\,x^4 +
		\frac{80}{3}\,x^5 
		+ \ln	x\Bigl( \frac{136}{3}\,x^3 - \frac{68}{3}\,x^4 +
			\frac{32}{5}\,x^5 \Bigr)\biggr] 
		\\&
		+ \tau^3 \, \biggl[ 304\,x^4 - \frac{1744}{3}\,x^5 +
		\frac{5504}{15}\,x^6 - \frac{448}{5}\,x^7
		\\&\mbox{\hspace{3em}}
		+ \ln x\Bigl(\frac{592}{3}\,x^4 - \frac{2128}{15}\,x^5 +
			\frac{1184}{15}\,x^6 - \frac{128}{7}\,x^7\Bigr) \biggr]
		\\&
		+ \tau^4 \, \biggl[ \frac{4480}{3}\,x^5 -
		\frac{156256}{45}\,x^6 +
		\frac{16192}{5}\,x^7 - \frac{164704}{105}\,x^8 +
		\frac{97408}{315}\,x^9 
		\\&\mbox{\hspace{3em}}
		+ \ln x\Bigl(\frac{12608}{15}\,x^5 - \frac{3904}{5}\,x^6
		+ \frac{67712}{105}\,x^7 - \frac{2080}{7}\,x^8 + 
		\frac{512}{9}\,x^9\Bigr) \biggr]\Biggr\}
		+ \ldots\,,
	\end{split}
\end{equation}
such that
\begin{equation}
	\begin{split}
		I_5 &= 
		- \frac{1}{2} 
		- \frac{5}{18}\tau 
		- \frac{29}{150}\tau^{2} 
		- \frac{4882}{33075}\tau^{3} 
		- \frac{11786}{99225}\tau^{4} 
		- \frac{3564004}{36018675}\tau^{5} 
		\\&
		- \frac{95238032}{1127251125}\tau^{6} 
		- \frac{745588736}{10145260125}\tau^{7} 
		- \frac{190175733376}{2931980176125}\tau^{8} 
		+ \ldots\,.
	\end{split}
\end{equation}
The remaining integrals in Ref.\,\cite{Spira:1995rr} are more complex,
but the integration of their expansion in $\tau$ can always be evaluated
in an elementary way.

We note in passing that for the coefficient of $\tau^{100}$, as it is
required by our general ansatz, the integers in the numerator and the
denominator are roughly of order $10^{180}$; this should give an
idea of the intermediate expressions' complexity.

Nevertheless, this expression, together with the corresponding expansion
of the ansatz, can be fed into {\code Mathematica} in order to solve the
resulting system of linear equations. One finds
\begin{equation}
	\begin{split}  
		I_5 &= 
		\frac{ \theta }{{\left(1-\theta\right)}^2}\biggl[
			4\Li_3(\theta)
			+8\Li_3(-\theta)
			-3\Li_2(\theta)\ln\theta
			-4\Li_2(-\theta)\ln\theta
			\\&\qquad
			-\ln(1-\theta){\ln^2\theta}
			+\zeta_2\ln\theta
			+ 2\zeta_3
			\biggr]
		+\frac{{\theta}^2 }{2
		{\left(1-\theta\right)}^3}{\ln^3\theta}.
	\end{split}
\end{equation}

Needless to say that the result obtained for $C_1^H$ in this way is in
agreement with \eqn{eq::c1h}, thus proving the consistency of the
analytical result of Ref.\,\cite{Fleischer:2004vb} and the integral
representation of Ref.\,\cite{Spira:1995rr}.

Let us add a few more remarks concerning the construction of the system
of linear equations. It happens that the structure of the ansatz can be
restricted already from general considerations:
The expansions of the entities calculated in this paper all consist
only of integer powers of $\tau$ multiplied by rational coefficients.
The expansions of Harmonic Polylogarithms of argument $\theta$, on the
other hand, contain noninteger powers of $\tau$, irrational numbers like
$\zeta_n$, logarithms of $\tau$, and have a non-vanishing imaginary
part.  The fact that such terms do not appear in the expansions of the
integrals to be matched produces a lot of equations that constrain our
ansatz, regardless of the specific integral to be calculated.

\subsection[Decay rate $A\to \gamma\gamma$]{%
  Decay rate \bld{A\to \gamma\gamma}}

After confirming the analytical result for $\Gamma(H\to \gamma\gamma)$
through \nlo{}, we are now ready to apply the method to the pseudo-scalar
case. Assuming the Minimal Supersymmetric Standard Model (\mssm{}) as
underlying theory, we write, in analogy to \eqn{eq::hgamgam}:
\begin{equation}\label{def:gammaa}
	\Gamma(A\to\gamma\gamma) = \frac{G_F \alpha^2 m_A^3}{32 \sqrt{2}
	\pi^3} 
	\abs{ \sum_l Q_l^2 g_l^A A_l^A(\tau_l) 
		+ 3\sum_q Q_q^2 g_q^A A_q^A(\tau_q) 
		+ \sum_{\tilde{\chi}^\pm} g_{\tilde{\chi}^\pm}^A 
		A_{\tilde{\chi}^\pm}^A(\tau_{\tilde{\chi}^\pm})},
\end{equation}
with the lepton ($l$) and chargino $(\chi^\pm)$ induced amplitudes
\begin{equation}
	A_l^{A}(\tau) 
	= A_{\tilde{\chi}^\pm}^{A}(\tau) 
	= \frac{f(\tau)}{\tau} \equiv F_0^A(\tau)\,,
\label{eq::f0a}
\end{equation}
where $f(\tau)$ has been defined in \eqn{eq:higgs:f}.  Note that there
is no contribution from the $W$ as loop particle due to CP
invariance. In \eqn{def:gammaa}, $m_A$ is the mass of the pseudo-scalar
Higgs, and the $\tau$-variables are defined according to \eqn{def:tau},
with $m_A$ instead of $m_H$.  The specific values of the couplings
$g_{l,q,\tilde{\chi}^\pm}^A$ are irrelevant for our analysis; they can be
found in Ref.\,\cite{Spira:1995rr}.

In analogy to \eqn{eq::aqh}, the quark-induced amplitude is written as
\begin{equation}
	A_q^{A}(\tau) = F_0^A(\tau) \biggl[
	1 + \api \bigl(
		C_1^{A}(\tau) + C_2^{A}(\tau) \ln\frac{4\tau\mu^2}{m_A^2}
	\bigr)\biggr],
\end{equation}
where again $C_2^{A}$ can be derived through a renormalization group
equation analogous to \eqn{eq::rge}:
\begin{equation}
\begin{split}
F_0^A\,C_2^A &= \frac{2}{\tau}\left[
f(\tau) - \tau\, f'(\tau)\right]\,.
\label{eq::c2a}
\end{split}
\end{equation}

Remarkably, when using the same ansatz as in the scalar case of
\sct{sec::hgamgam}, the resulting system of linear equations has no
solution. The necessary generalization is to allow for terms $\sim
(1+\theta)^{-1}$ multiplying the \hpl{}s, reflecting the well-known
threshold singularity in the pseudo-scalar case at $m_A=2m_q$.  Once
this is done, the \ibe{} approach yields
\begin{equation}
  \begin{split}
    F_0^A\,& C_1^A =\\&
    -\frac{ \theta \left(1+\theta^2\right) }
    {{\left(1-\theta\right)}^3\left(1+\theta\right)}
    \biggl[
       72\Li_4(\theta)
      + 96\Li_4(-\theta)
      -\frac{128}{3}\Big[\Li_3(\theta) + \Li_3(-\theta)\Big]\ln \theta
      \\&\qquad
      +\frac{28}{3}\Li_2(\theta) \ln^2 \theta
      +\frac{16}{3}\Li_2(-\theta) \ln^2 \theta
      +\frac{1}{18} \ln^4 \theta
      \\&\qquad
      +\frac{8}{3}\zeta_2 \ln^2 \theta
      +\frac{32}{3}\zeta_3\ln \theta
      +12\zeta_4
      \biggr]
    \\&
    +\frac{ \theta }{ {\left(1-\theta\right)}^2}
    \biggl[
      -\frac{56}{3}\Li_3(\theta)
      -\frac{64}{3}\Li_3(-\theta)
      +16\Li_2(\theta)\ln \theta
      \\&\qquad
      +\frac{32}{3}\Li_2(-\theta)\ln \theta
      +\frac{20}{3}\ln(1-\theta)\ln^2 \theta
      -\frac{8}{3}\zeta_2\ln \theta
      +\frac{8}{3}\zeta_3
      \biggr]
    \\&
    +\frac{ 2\,\theta \left(1+\theta\right) }{
      {3\left(1-\theta\right)}^3}\, \ln^3 \theta
    \,.
  \end{split}
\end{equation}

\section{Virtual corrections for \bld{gg\to H/A}}

An interesting application of our method is the analytical evaluation of
the virtual two-loop corrections for Higgs production in gluon fusion
for arbitrary values of the quark and Higgs boson mass.

Following Ref.\,\cite{Spira:1995rr}, we write the inclusive \nlo{} cross
section as
\begin{equation}
  \begin{split}
    \sigma(pp\to \Phi+X) &= \sigma_0^\Phi\left[1 +
      C^\Phi\,\api\right]\tau_\Phi\frac{\dd{\cal L}^{gg}}{\dd\tau_\Phi} 
    + \Delta\sigma^\Phi_{gg}
    + \Delta\sigma^\Phi_{gq}
    + \Delta\sigma^\Phi_{q\bar q}\,,\\
    \qquad \Phi&\in\{H,A\}\,,
    \label{eq::sigma}
  \end{split}
\end{equation}
where $\tau_\Phi = m_\Phi^2/s$ with the center-of-mass energy $s$, and
\begin{equation}
\begin{split}
\frac{\dd{\cal L}^{gg}}{\dd\tau} &= \int_\tau^1\frac{\dd x}{x}
g(x,\mu_F)\,g(\tau/x,\mu_F)\,,
\end{split}
\end{equation}
with the gluon density functions $g(x,\mu_F)$, depending on the
factorization scale $\mu_F$. The normalization factors are
\begin{equation}
\begin{split}
\sigma_0^H &= \frac{G_F\alpha_s^2}{288\sqrt{2}\pi}
\left|\sum_q g_q^H F_0^H(\tau_q)\right|\,,\qquad
\sigma_0^A = \frac{G_F\alpha_s^2}{128\sqrt{2}\pi}
\left|\sum_q g_q^A F_0^A(\tau_q)\right|\,,
\end{split}
\end{equation}
with $F_0^{H/A}$ defined in Eqs.\,(\ref{eq::alhawh}), (\ref{eq::f0a}).
$C^\Phi$ denotes the contributions from the virtual two-loop
corrections, regularized by the infrared singular part of the cross
section for real gluon emission (see Ref.\,\cite{Spira:1995rr} for details).
It can be decomposed into
\begin{equation}
\begin{split}
C^\Phi &= \pi^2 + c^\Phi +
2\,\beta_0\,\ln\frac{\mu^2}{m_\Phi^2}\,,
\label{eq::cphi}
\end{split}
\end{equation}
where $\beta_0 = 11/4-n_f/6$ is the lowest-order $\beta$ function of
\qcd{} for $n_f$ active quark flavors. The $\Delta\sigma^\Phi_{ij}$
denote the contributions from radiation of quarks and gluons with
initial state partons $i,j\in\{q,\bar q,g\}$. At \nlo{} perturbation
theory, they correspond to massive one-loop three- and four-point
functions which can be evaluated analytically using standard
techniques~\cite{Passarino:1978jh} (see also Ref.\,\cite{Denner:1991kt}).
They shall not be considered any further in this paper.

The coefficient $c^\Phi$ of the virtual corrections in \eqn{eq::cphi}
is parameterized as
\begin{equation}
\begin{split}
c^\Phi &= \Re\left\{
\frac{\sum_q g_q^\Phi F_0^\Phi(\tau_q)\left( B_1^\Phi(\tau_q) 
  + B_2^\Phi(\tau_q)\,\ln\frac{\mu^2}{m_q^2} \right)}{%
  \sum_q g_q^\Phi F_0^\Phi(\tau_q)}
\right\}\,.
\end{split}
\end{equation}
Similar to the decay rates, $B_2^\Phi$ follows from renormalization group
considerations:
\begin{equation}
\begin{split}
B_2^\Phi(\tau) &= 2\,C_2^\Phi(\tau)\,,\qquad \Phi\in\{H,A\}\,,
\end{split}
\end{equation}
with $C_2^\Phi$ from Eqs.\,(\ref{eq::c2h}) and (\ref{eq::c2a}). The
factor of 2 arises from the fact that $C^\Phi$ in \eqn{eq::sigma} is
defined at the level of the cross section rather than the amplitude.

$B_1^\Phi$ is known again in terms of one-dimensional integrals, filling
several pages ($I_1,\ldots, I_8$ in App.\,A,B of
Ref.\,\cite{Spira:1995rr}).  Following the method described in
\sct{sec::method}, we expanded the integrands for $B_1^\Phi(\tau)$
around $\tau=0$ up to order $\tau^{100}$ and mapped them onto a set of
suitably chosen basis functions.

\FIGURE{%
    \begin{tabular}{cc}
      \includegraphics[width=.3\textwidth]{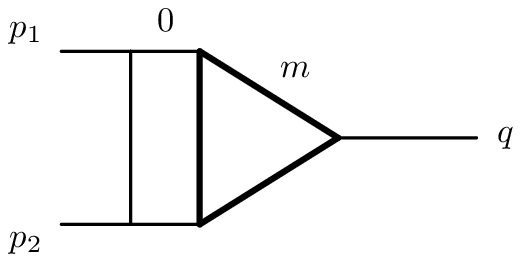}\qquad &\qquad
      \raisebox{.3em}{\includegraphics[width=.3\textwidth]{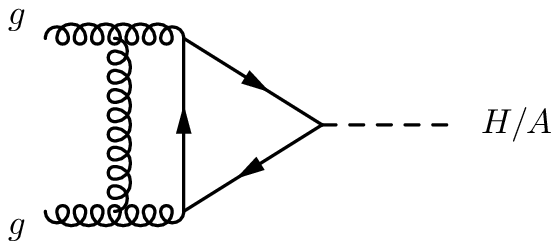}}
      \\[0em]
      (a) & (b)
    \end{tabular}
    \parbox{.9\textwidth}{
      \caption[]{\sloppy
	(a) Feynman integral contributing to the production rate $gg\to
	H/A$ but not to the decay rate $H/A\to\gamma\gamma$; it arises
	due to the self-coupling of gluons, see (b).
	\label{fig::p126}
        }}
}

In the case of gluonic Higgs production, a new class of integrals occurs
that cannot be expressed in terms of the functions used for the decay
rates. The corresponding scalar diagram is shown in
\fig{fig::p126}\,(a); it arises from the physical process due to the
self-interaction of gluons, see \fig{fig::p126}\,(b).  The scalar
integral has been evaluated in Ref.\,\cite{Davydychev:2003mv}. The
result contains a Harmonic Polylogarithm of weight four with two indices
different from zero.  We therefore enlarge our basis to include this
kind of functions. Apart from that, the method works exactly like in the
case of the decay rates, described in \sct{sec::method}.

With $\theta$ defined in \eqn{eq::thetadef}, we find for the scalar case
\begin{equation}
  \begin{split}
    F_0^H\,&B_1^H =\\&
    \frac{ \theta {\left(1+\theta\right)}^2 }{{\left(1-\theta\right)}^4}
    \biggl[ 72	\HPL(1,0,-1,0;\theta) 
      + 6\ln(1-\theta)\ln^3\theta
      - 36\,\zeta_2\Li_2(\theta) 	
      \\&\qquad
      - 36\,\zeta_2\ln(1-\theta)\ln\theta    
      -108\,\zeta_3\ln(1-\theta) 
      \\&\qquad
      - 64\Li_3(-\theta)
      + 32\Li_2(-\theta)\ln\theta
      - 8\,\zeta_2\ln\theta 
      \biggr]
      \\&
    -\frac{ 36\,\theta \left(5+5\,\theta+11\,{\theta}^2+11\,{\theta}^3\right)
    }{{\left(1-\theta\right)}^5}\Li_4(-\theta) 
    -\frac{ 36\,\theta \left(5+5\,\theta+7\,{\theta}^2+7\,{\theta}^3\right)
    }{{\left(1-\theta\right)}^5}\Li_4(\theta) 
    \\&
    +\frac{ 4\,\theta \left(1+\theta\right) \left(23+41\,
    {\theta}^2\right)}{{\left(1-\theta\right)}^5}  
    \biggl[
      \Li_3(\theta)
      +\Li_3(-\theta)
      \biggr] \ln\theta 
    \\&
    -\frac{ 16\,\theta \left(1+\theta+{\theta}^2+{\theta}^3\right)
    }{{\left(1-\theta\right)}^5}\Li_2(-\theta)\ln^2\theta 
    -\frac{ 2\,\theta \left(5+5\,\theta+23\,{\theta}^2+23\,{\theta}^3\right)
    }{{\left(1-\theta\right)}^5}\Li_2(\theta)\ln^2\theta 
    \\&
    +\frac{ \theta \left(5+5\,\theta-13\,{\theta}^2-13\,{\theta}^3\right)
    }{24 {\left(1-\theta\right)}^5}\ln^4\theta 
    +\frac{ \theta \left(1+\theta-17\,{\theta}^2-17\,{\theta}^3\right)
    }{{\left(1-\theta\right)}^5}\zeta_2\ln^2\theta 
    \\&
    +\frac{ 2\,\theta \left(11+11\,\theta-43\,{\theta}^2-43
    {\theta}^3\right) }{{\left(1-\theta\right)}^5}\zeta_3\ln\theta 
    + \frac{ 36\,\theta \left(1+\theta-3\,{\theta}^2-3\,{\theta}^3\right)
    }{{\left(1-\theta\right)}^5}\zeta_4
    \\&
    -\frac{ 2\,\theta \left(55+82\,\theta+55\,{\theta}^2\right)
    }{{\left(1-\theta\right)}^4}\Li_3(\theta) 
    +\frac{ 2\,\theta \left(51+74\,\theta+51\,{\theta}^2\right)
    }{{\left(1-\theta\right)}^4}\Li_2(\theta)\ln\theta 
    \\&
    +\frac{ \theta \left(47+66\,\theta+47\,{\theta}^2\right)
    }{{\left(1-\theta\right)}^4}\ln(1-\theta)\ln^2\theta 
    +\frac{ \theta \left(6+59\,\theta+58\,{\theta}^2+33\,{\theta}^3\right)
    }{3 {\left(1-\theta\right)}^5}\ln^3\theta 
    \\&
    +\frac{ 2\,\theta \left(31+34\,\theta+31\,{\theta}^2\right)
    }{{\left(1-\theta\right)}^4}\zeta_3
    +\frac{ 3\,\theta \left(3+22\,\theta+3\,{\theta}^2\right) }{
    2{\left(1-\theta\right)}^4}\ln^2\theta
    \\&
    -\frac{ 24\,\theta \left(1+\theta\right)
    }{{\left(1-\theta\right)}^3}\ln\theta
    -\frac{ 94\,\theta}{{\left(1-\theta\right)}^2}\,.
  \end{split}
\end{equation}

For the pseudo-scalar case we get
\begin{equation}
  \begin{split}
    F_0^A\,&B_1^A =\\&
    \frac{ \theta }{ {\left(1-\theta\right)}^2}\biggl[ 
      48\HPL(1,0,-1,0;\theta)
      + 4\ln(1-\theta)\ln^3\theta
      - 24\,\zeta_2 \Li_2(\theta)
      \\&\qquad
      - 24\,\zeta_2 \ln(1-\theta)\ln\theta
      - 72\,\zeta_3 \ln(1-\theta)
      - \frac{220}{3}\Li_3(\theta)
      - \frac{128}{3}\Li_3(-\theta)
      \\&\qquad
      + 68\Li_2(\theta)\ln\theta
      + \frac{64}{3}\Li_2(-\theta)\ln\theta
      + \frac{94}{3}\ln(1-\theta)\ln^2\theta
      \\&\qquad
      - \frac{16}{3}\zeta_2\ln\theta
      + \frac{124}{3}\zeta_3
      + 3\ln^2\theta 
      \biggr]
    \\& 
    - \frac{ 24\,\theta \left(5+7\,{\theta}^2\right)
    }{{\left(1-\theta\right)}^3 \left(1+\theta\right)}\Li_4(\theta) 
    -\frac{ 24\,\theta \left(5+11\,{\theta}^2\right)
    }{{\left(1-\theta\right)}^3 \left(1+\theta\right)}\Li_4(-\theta) 
    \\&
    +\frac{ 8\,\theta \left(23+41\,{\theta}^2\right) }
    {3{\left(1-\theta\right)}^3 \left(1+\theta\right)}
    \biggl[
      \Li_3(\theta)
      +\Li_3(-\theta)
      \biggr] \ln\theta
    -\frac{ 4\,\theta \left(5+23\,{\theta}^2\right) }{
    3{\left(1-\theta\right)}^3
    \left(1+\theta\right)}\Li_2(\theta)\ln^2\theta 
    \\&
    -\frac{ 32\,\theta \left(1+{\theta}^2\right) }{
    3{\left(1-\theta\right)}^3
    \left(1+\theta\right)}\Li_2(-\theta)\ln^2\theta 
    +\frac{ \theta \left(5-13\,{\theta}^2\right) }{36
    {\left(1-\theta\right)}^3 \left(1+\theta\right)}\ln^4\theta 
    \\& 
    +\frac{ 2\,\theta \left(1-17\,{\theta}^2\right) }{
    3{\left(1-\theta\right)}^3 \left(1+\theta\right)}\zeta_2\ln^2\theta 
    +\frac{ 4\,\theta \left(11-43\,{\theta}^2\right) }{
    3{\left(1-\theta\right)}^3 \left(1+\theta\right)}\zeta_3\ln\theta 
    \\& 
    +\frac{ 24\,\theta \left(1-3\,{\theta}^2\right)
    }{{\left(1-\theta\right)}^3 \left(1+\theta\right)}\zeta_4 
    +\frac{ 2\,\theta \left(2+11\,\theta\right) }{
    3{\left(1-\theta\right)}^3}\ln^3\theta \,.
  \end{split}
\end{equation}

One may notice that the Harmonic Polylogarithm appearing in
Ref.\,\cite{Davydychev:2003mv} is different from the one contributing to
$B_1^H$ and $B_1^A$.  However, this is only due to an arbitrariness when
choosing a basis of Harmonic Polylogarithms.  In fact,
\begin{equation}
	\begin{split}
		8\HPL(1,0,-1,0;\theta) &= 
		-8\HPL(-1,0,0,1;-\theta) 
		- 2\Snp_{2,2}\bigl(\theta^2\bigr) 
		+	8\Snp_{2,2}(\theta) 
		+	8\Snp_{2,2}(-\theta) 
		\\&
		+ 4\ln \theta \Snp_{1,2}\bigl(\theta^2\bigr) 
		- 8\ln \theta \Snp_{1,2}(\theta) 
		- 8\ln \theta \Snp_{1,2}(-\theta)
		\\&
		- 8\ln(1-\theta) \Li_3(-\theta)
		+ 8\ln(1-\theta) \ln \theta \Li_2(-\theta),
	\end{split}
\end{equation}
as can be seen using Appendix B of Ref.\,\cite{Czakon:2004wm}, for
example. Using $\HPL(1,0,-1,0;\theta)$, all $\Snp_{2,2}$ and
$\Snp_{1,2}$ cancel in our final result.

\section{Numerical Results}

As already mentioned in \sct{sec::method}, the comparison of the
final result with a numerical evaluation of the original integral
provides one of the most important checks of the calculation. 

\FIGURE{%
    \begin{tabular}{c}
    \includegraphics[width=.8\textwidth]{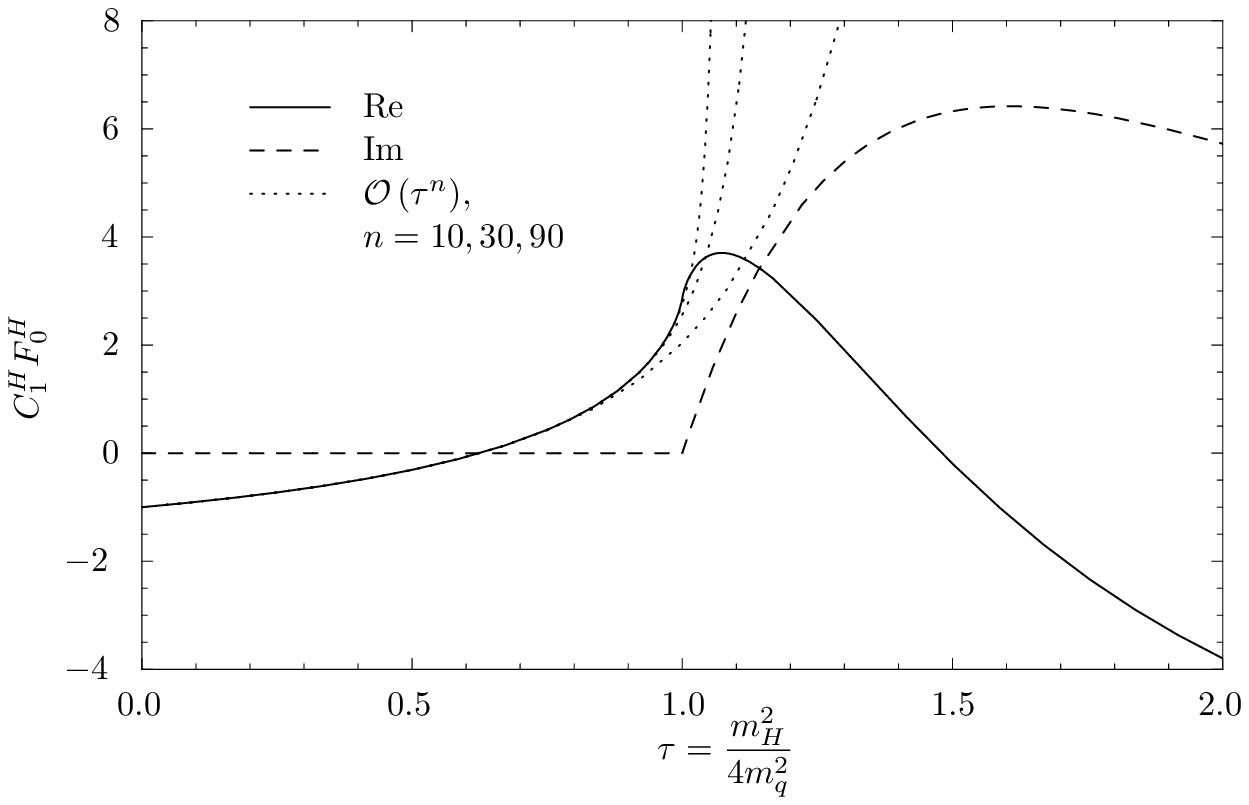}\\
    (a)\\[1em]
    \includegraphics[width=.8\textwidth]{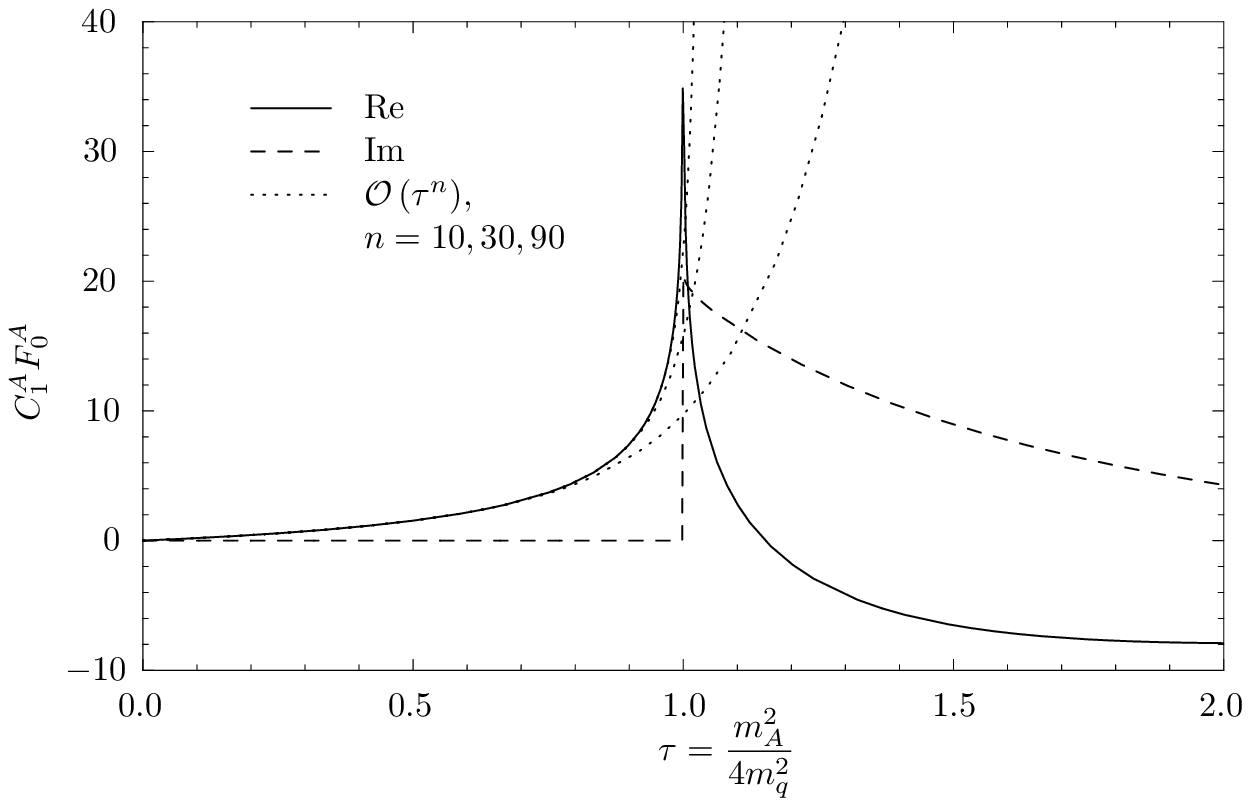}\\
    (b)
    \end{tabular}
    \caption[]{ Two-loop contributions to the (a) scalar and (b)
      pseudo-scalar Higgs decay rate into photons. The solid and dashed
      lines show the real and the imaginary part, respectively. The
      dotted lines correspond to the power series expansion up to order
      $\tau^n$ with $n=10,30,90$.
    \label{fig::c1}}
}

\FIGURE{%
    \begin{tabular}{c}
    \includegraphics[width=.8\textwidth]{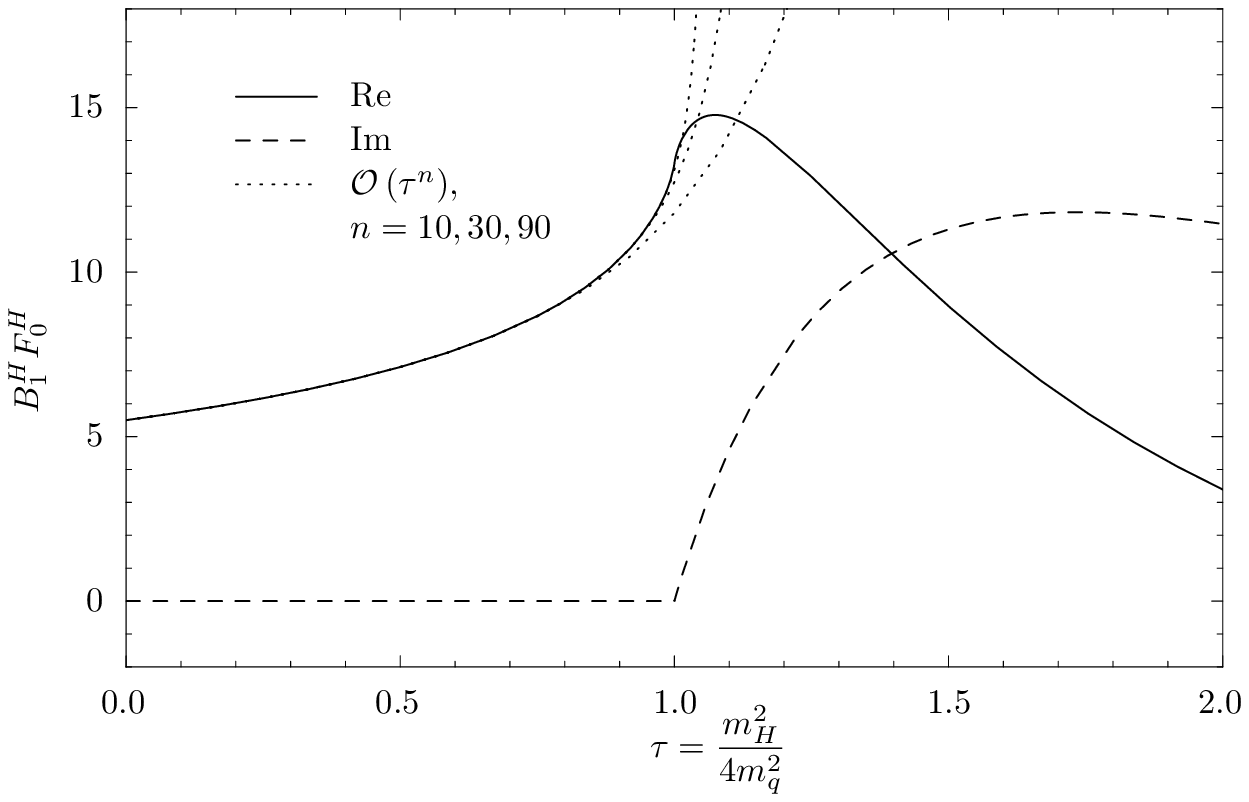}\\
    (a) \\[1em]
    \includegraphics[width=.8\textwidth]{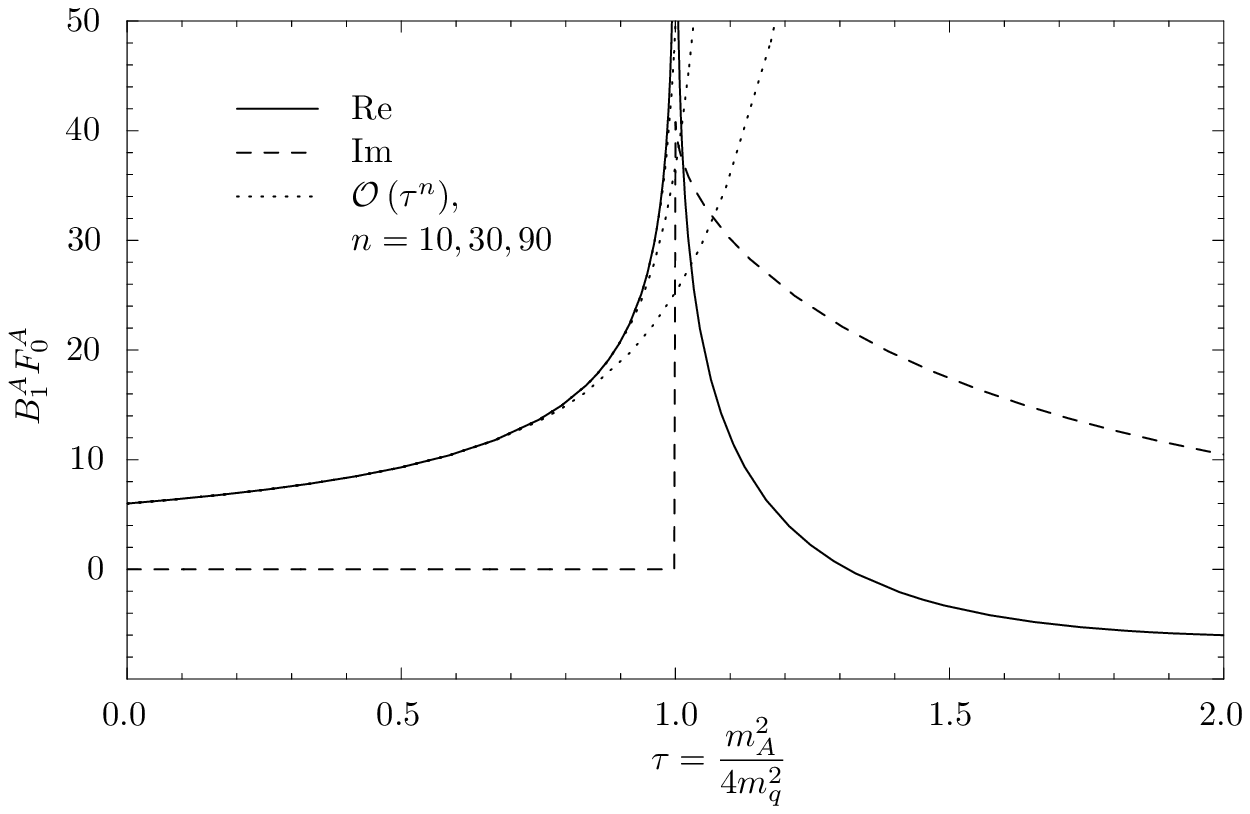}\\
    (b)
    \end{tabular}
    \caption[]{ Infrared regularized virtual two-loop corrections to
      the (a) scalar and (b) pseudo-scalar Higgs production rate through
      gluon fusion.  The solid and dashed lines show the real and the
      imaginary part, respectively. The dotted lines correspond to the
      power series expansion up to order $\tau^n$ with $n=10,30,90$.
      \label{fig::b1}}
}

The solid and dashed lines of \fig{fig::c1}\,(a) and (b) show the real
and imaginary part of $C_1^HF_0^H$ and $C_1^AF_0^A$, respectively.  The
dotted lines in \fig{fig::c1} show the results obtained from the
intermediate power series when keeping terms of order $\tau^{n}$ with
$n=10,30,90$. Note that, as expected, the power series does not converge
towards the analytic result beyond the radius of convergence, given by
$\tau=1$. In particular, the imaginary part is always zero. Thus, for
$\tau\geq 1$, the result arises solely from analytic continuation of the
terms reconstructed through \ibe{}.

In addition, we were able to reproduce Figs.\,5 and~18 of
Ref.\,\cite{Spira:1995rr}, using our results, to perfect agreement.  We
find a similar picture for the virtual corrections to gluon fusion,
shown in \fig{fig::b1}\,(a) and (b) for the scalar and the pseudo-scalar
case, respectively. The numerical evaluation of $\HPL(1,0,-1,0;\theta)$
was done with the help of the {\code Mathematica} file {\tt HPL4.m} in
\cite{hpl4}.

\section{Conclusions}

The two-loop \qcd{} results for the decay rate of a scalar or
pseudo-scalar Higgs boson into photons, $H/A\to \gamma\gamma$, as well
as for the virtual corrections to the production modes $gg\to H/A$ were
presented in closed analytical form. In order to obtain these result, we
first expanded the known one-dimensional integral representations in
terms of small Higgs masses, and subsequently mapped this expansion onto
a set of analytic functions. The final results, both for their real and
imaginary part, are valid for arbitrary values of the quark and Higgs
boson mass. They contain only polylogarithms or simpler functions and,
in the case of $gg\to H/A$, one Harmonic Polylogarithm.

Our formulas should be useful for implementations into physics analysis
programs, or for quickly obtaining analytical limits to arbitrary accuracy.

Let us finish by pointing out that the \ibe{} method, in various
flavors, has been quite useful already in the past (see
Refs.\,\cite{Czarnecki:1995jt,Fleischer:1998nb,vanRitbergen:1999fi,
Baikov:2001aa,Harlander:2002vv,Kilgore:2002sk},
for example). Its combination with asymptotic expansions may even carry
the potential for an algorithmic evaluation of Feynman integrals.
However, this not only requires much more efficient computer algebra
tools for the expansion of Feynman diagrams. The more important task is
to find suitable bases for certain classes of Feynman integrals.  We
believe that this is certainly a task worth pursuing.

\paragraph{Acknowledgments.}

We are indebted to P.~Baikov and A.~Grozin for valuable advice
concerning {\code Form}, {\code Reduce} and the {\code Taylor} package,
to O.~Tarasov for encouragement and useful conversations, to M.~Spira
for communications concerning Ref.\,\cite{Spira:1995rr} and for
providing us with an unpublished {\tt FORTRAN} routine for numerical
checks, and to O.~Veretin for sharing his experience with Harmonic
Polylogarithms and the \ibe{} method. Special thanks go to J.H.~K\"uhn
for encouragement and advice.

We kindly acknowledge financial support by {\it Deutsche
  Forschungsgemeinschaft}
(contract HA\,2990/2-1, {\it Emmy Noether program}).

\def\app#1#2#3{{\it Act.~Phys.~Pol.~}\jref{\bf B #1}{#2}{#3}}
\def\apa#1#2#3{{\it Act.~Phys.~Austr.~}\jref{\bf#1}{#2}{#3}}
\def\annphys#1#2#3{{\it Ann.~Phys.~}\jref{\bf #1}{#2}{#3}}
\def\cmp#1#2#3{{\it Comm.~Math.~Phys.~}\jref{\bf #1}{#2}{#3}}
\def\cpc#1#2#3{{\it Comp.~Phys.~Commun.~}\jref{\bf #1}{#2}{#3}}
\def\epjc#1#2#3{{\it Eur.\ Phys.\ J.\ }\jref{\bf C #1}{#2}{#3}}
\def\fortp#1#2#3{{\it Fortschr.~Phys.~}\jref{\bf#1}{#2}{#3}}
\def\ijmpc#1#2#3{{\it Int.~J.~Mod.~Phys.~}\jref{\bf C #1}{#2}{#3}}
\def\ijmpa#1#2#3{{\it Int.~J.~Mod.~Phys.~}\jref{\bf A #1}{#2}{#3}}
\def\jcp#1#2#3{{\it J.~Comp.~Phys.~}\jref{\bf #1}{#2}{#3}}
\def\jetp#1#2#3{{\it JETP~Lett.~}\jref{\bf #1}{#2}{#3}}
\def\jhep#1#2#3{{\small\it JHEP~}\jref{\bf #1}{#2}{#3}}
\def\mpl#1#2#3{{\it Mod.~Phys.~Lett.~}\jref{\bf A #1}{#2}{#3}}
\def\nima#1#2#3{{\it Nucl.~Inst.~Meth.~}\jref{\bf A #1}{#2}{#3}}
\def\npb#1#2#3{{\it Nucl.~Phys.~}\jref{\bf B #1}{#2}{#3}}
\def\nca#1#2#3{{\it Nuovo~Cim.~}\jref{\bf #1A}{#2}{#3}}
\def\plb#1#2#3{{\it Phys.~Lett.~}\jref{\bf B #1}{#2}{#3}}
\def\prc#1#2#3{{\it Phys.~Reports }\jref{\bf #1}{#2}{#3}}
\def\prd#1#2#3{{\it Phys.~Rev.~}\jref{\bf D #1}{#2}{#3}}
\def\pR#1#2#3{{\it Phys.~Rev.~}\jref{\bf #1}{#2}{#3}}
\def\prl#1#2#3{{\it Phys.~Rev.~Lett.~}\jref{\bf #1}{#2}{#3}}
\def\pr#1#2#3{{\it Phys.~Reports }\jref{\bf #1}{#2}{#3}}
\def\ptp#1#2#3{{\it Prog.~Theor.~Phys.~}\jref{\bf #1}{#2}{#3}}
\def\ppnp#1#2#3{{\it Prog.~Part.~Nucl.~Phys.~}\jref{\bf #1}{#2}{#3}}
\def\rmp#1#2#3{{\it Rev.~Mod.~Phys.~}\jref{\bf #1}{#2}{#3}}
\def\sovnp#1#2#3{{\it Sov.~J.~Nucl.~Phys.~}\jref{\bf #1}{#2}{#3}}
\def\sovus#1#2#3{{\it Sov.~Phys.~Usp.~}\jref{\bf #1}{#2}{#3}}
\def\tmf#1#2#3{{\it Teor.~Mat.~Fiz.~}\jref{\bf #1}{#2}{#3}}
\def\tmp#1#2#3{{\it Theor.~Math.~Phys.~}\jref{\bf #1}{#2}{#3}}
\def\yadfiz#1#2#3{{\it Yad.~Fiz.~}\jref{\bf #1}{#2}{#3}}
\def\zpc#1#2#3{{\it Z.~Phys.~}\jref{\bf C #1}{#2}{#3}}
\def\ibid#1#2#3{{ibid.~}\jref{\bf #1}{#2}{#3}}

\newcommand{\jref}[3]{{\bf #1}, #3 (#2)}
\newcommand{\bibentry}[4]{#1, #3.}
\newcommand{\arxiv}[1]{{\tt arXiv:#1}}

\end{document}